\documentclass[12pt]{article}
\usepackage[dvips]{color}
\usepackage{epsfig}
\usepackage{amsmath}
\usepackage{graphicx}
\def\Box{\hbox{$\rlap{$\sqcup$}\sqcap$}}

\begin{document}
\title{\bf Reconstructing a String-Inspired Non-minimally Coupled Quintom Model
\footnote{\bf Dedicated to the 60 year Jubilee of Professor Reza
Mansouri}}

\author{{M. R. Setare $^{a}$ \thanks{Email: rezakord@ipm.ir}\hspace{1mm}
, J. Sadeghi $^{b}$\thanks{Email: pouriya@ipm.ir}\hspace{1mm} and A. R. Amani$^{c}$ \thanks{Email: a.r.amani@iauamol.ac.ir}}\\
$^{a}${\small {\em Department of Science, Payame Noor University, Bijar, Iran }}\\
$^b${\small {\em Sciences Faculty, Department of Physics, Mazandaran
University,}}\\{\small {\em P .O .Box 47415-416, Babolsar, Iran}}\\
$^c$ {\small {\em  Department of Physics, Islamic Azad University - Ayatollah Amoli Branch,}}\\
        {\small {\em P.O.Box 678, Amol, Iran}}}
\maketitle

\begin{abstract}
\noindent Motivated by the recent work of Zhang and Chen \cite{bin},
we generalize their work to the non-minimally coupled case. We
consider a quintom model of dark energy with a single scalar field
$T$ given by a Lagrangian which inspired by tachyonic Lagrangian in
string theory. We consider non-minimal coupling of tachyon field to
the scalar curvature, then we reconstruct this model in the light of
three forms of parametrization for dynamical dark energy.

\end{abstract}
\newpage
\section{Introduction}
Nowadays it is plainly believed that the universe is experiencing an
accelerated expansion. Recent observations from type Ia supernovae
\cite{1} in associated with Large Scale Structure \cite{2} and
Cosmic Microwave Background anisotropies \cite{3} have provided main
evidence for this cosmic acceleration. In order to explain why the
cosmic acceleration happens, many theories have been proposed.
Although theories of trying to modify Einstein equations constitute
a big part of these attempts, the mainstream explanation for this
problem, however, is known as theories of dark energy.\\
The combined analysis of cosmological observations suggests that the
universe consists of about $70\%$ dark energy, $30\%$ dust matter
(cold dark matter plus baryons), and negligible radiation. Although
the nature and origin of dark energy are unknown, we still can
propose some candidates to describe it, namely  since we do not know
where this dark energy comes from, and how to compute it from the
first principles, we search for phenomenological models. The
astronomical observations will then select one of these models. The
most obvious theoretical candidate of dark energy is the
cosmological constant $\lambda$ (or vacuum energy)
\cite{Einstein:1917,cc} which has the equation of state parameter
$w=-1$. However, as it is well known, there are two difficulties
that arise from the cosmological constant scenario, namely the two
famous cosmological constant problems --- the ``fine-tuning''
problem and the ``cosmic coincidence'' problem \cite{coincidence}.
An alternative proposal for dark energy is the dynamical dark energy
scenario. This  dynamical proposal is often realized by some scalar
field mechanism which suggests that the specific energy form with
negative pressure is provided by a scalar field evolving down a
proper potential. Primary scalar field candidate for dark energy was
quintessence scenario \cite{{rat},{zlat}}, a fluid with the
parameter of the equation of state lying in the range, $-1< \omega<
{-1 \over 3}$. The analysis of the properties of dark energy from
recent observations mildly favor models with $w$ crossing -1 in the
near past.\\
Meanwhile for the phantom model\cite{pha} of dark energy which has
the opposite sign of the kinetic term compared with the quintessence
in the Lagrangian, one always has $\omega\leq -1$. Neither the
quintessence nor the phantom alone can fulfill the transition from
$\omega>-1$ to $\omega<-1$ and vice versa. But one can show
\cite{{c9},{c12},{c14},{c15}} that considering the combination of
quintessence and phantom in a joint model, the transition can be
fulfilled. This model, dubbed quintom, can produce a better fit to
the data than
more familiar models with $w\geq-1$.\\
To realize a viable quintom scenario of dark energy it needs to
introduce extra degree of freedom to the conventional theory with a
single fluid or a single scalar field. The first model of quintom
scenario of dark energy is given by Ref.\cite{c9} with two scalar
fields. This model has been studied in detail later on
\cite{{c12},{c14},{c15}} (to see the bouncing solution in the
universe dominated by quintom matter refer to \cite{bou}). Recently
there has been an upsurge in activity for constructing such model in
string theory \cite{c16}. In the context of string theory, the
tachyon field in the world volume theory of the open string
stretched between a D-brane and an anti-D-brane or a non-BPS D-brane
plays the role of scalar field in the quintom model \cite{c17}. The
effective action used in the study of tachyon cosmology consists of
the standard Einstein-Hilbert action and an effective action for the
tachyon field on unstable D-brane or D-brane anti D-brane system.
What distinguishes the tachyon action from the standard Klein-
Gordon form for scalar field is that the tachyon action is
non-standard and is of the " Dirac-Born-Infeld " form \cite{c18}.
The tachyon potential is derived from string theory itself and has
to satisfy some definite properties to describe
tachyon condensation and other requirements in string theory\cite{c20}.\\

\section{Reconstruction of non-minimally coupled tachyon gravity with extra term}
We consider the action Ref.\cite{c20} for tachyon non-minimally
coupled to gravity, then we add an extra term $T\Box {T}$ to the
usual terms in the square root of this action. In that case the
following action is the same as Ref.\cite{c21} just different to the
$ Rf(T),$
\begin{equation}
S=\int d^{4}x \sqrt{-g} \left[\frac{M_{P}^{2}}{2}Rf(T) -
AV(T)\sqrt{1-\alpha'g^{\mu\nu}\partial_{\mu}T\partial_{\nu}T+\beta'T
\Box T}\right],
\end{equation}
where $f(T)$ is a function of the tachyon $T$ and corresponds to the
non-minimal coupling factor. Here $V(T)$ is the tachyon potential
which is bounded and reaching its minimum asymptotically.
$M_{P}=\frac{1}{\sqrt{8\pi G}}$ is reduced
Planck mass.\\
The action (1) can be brought to the simpler form to derive the
equation of motion, energy density and pressure, by performing a
conformal transformation as follows:
\begin{eqnarray}
g_{\mu\nu}\longrightarrow f(T) g_{\mu\nu}.
\end{eqnarray}
The above conformal transformation yields to the following action:\\
$$S=\int d^{4}x\sqrt{-g}\
\Bigg[\frac{M_{P}^{2}}{2}(R-\frac{3}{2}\frac{f'^{2}}{f^{2}}\partial_{\mu}T\partial^{\mu}T)$$

\begin{eqnarray}
-A\tilde{V}(T)\sqrt{1-(\alpha'f(T)-2\beta'f'(T)T)\partial_{\mu}T\partial^{\mu}T+\beta'f(T)T
\Box T}\Bigg]\,
\end{eqnarray}
where $\tilde{V}(T)=\frac{V(T)}{f^{2}}$ is now the effective
potential of the tachyon.\\
 For a flat Friedman- Robertson- Walker
(FRW) universe and a homogenous scalar field $T$, we have

\begin{eqnarray}\label{eq4}
\ddot{T}+3H\dot{T}=\frac{2\left[(\frac{ff''+\beta'f'}{f^{2}})T\dot{T}^{2}-2
(\alpha'-2\beta'\frac{f'}{f}T)H\dot{T}\right]}{1+\frac{2\alpha'}{\beta'}-3\frac{f'}{f}T-
\frac{3M_{P}^{2}}{2}(\frac{f'}{f})^{2}\frac{T}{\psi}}=\gamma,
\end{eqnarray}
where
\begin{eqnarray}\label{pesi}\psi=\frac{\partial  \mathcal{L}}{\partial
\Box T}=-\frac{A\beta'\tilde{V}f T}{2h} \hspace{0.5cm}
h=-\frac{A\beta'\tilde{V}f T}{2\psi}
\end{eqnarray}
also we have  $$
h=\sqrt{1-(\alpha'f-2\beta'f'T)\partial_{\mu}T\partial^{\mu}T+\beta'fT
\Box T}$$ and $H=\frac{\dot{a}}{a}$ is
the Hubble parameter.\\
The energy momentum tensor $T^{\mu\nu}$ is given by the standard
definition: $$\delta_{g_{\mu\nu}}S=-\int
d^{4}x\frac{\sqrt{-g}}{2}T^{\mu\nu}\delta g_{\mu\nu}.$$ So the
energy density, and pressure are found to be
\begin{eqnarray}\label{rho}
\rho=A\tilde{V}h+\frac{d}{a^{3}dt}(a^{3}\psi\dot{T})+(\alpha'f-2\beta'f'T)
\frac{A\tilde{V}}{h}\dot{T}^{2}-2\dot{\psi}\dot{T}+\frac{3M_{P}^{2}}{4}(\frac{f'}{f}^{2})\dot{T}^{2},
\end{eqnarray}
\begin{eqnarray}
p=-A\tilde{V}h-\frac{d}{a^{3}dt}(a^{3}\psi\dot{T})+\frac{3M_{P}^{2}}{4}(\frac{f'}{f}^{2})\dot{T}^{2},
\end{eqnarray}
From equations (6) and (7) one can obtain the following expressions,
\begin{eqnarray}\label{plus1}
\rho+p=\frac{3M_{P}^{2}}{2}(\frac{f'}{f}^{2})\dot{T}^{2}+(\alpha'f-2\beta'f'T)
\frac{\tilde{V}A}{h}\dot{T}^{2}-2\dot{\psi}\dot{T},
\end{eqnarray}
By substituting $h$ from (\ref{pesi}) into the Eqs.(\ref{plus1}),
(\ref{rho})respectively we obtain
\begin{eqnarray}\label{plus2}
\rho+p=\frac{3M_{P}^{2}}{2}(\frac{f'}{f}^{2})\dot{T}^{2}-(\alpha'f-2\beta'f'T)
\frac{2\psi}{\beta'fT}\dot{T}^{2}-2\dot{\psi}\dot{T}=2\hat{K},
\end{eqnarray}
\begin{eqnarray}\label{rho2}
\rho=-(\alpha'f-2\beta'f'T)\frac{2\psi}{\beta'fT}\dot{T}^{2}
-\dot{\psi}\dot{T}+\frac{3M_{P}^{2}}{4}(\frac{f'}{f}^{2})\dot{T}^{2}-
\frac{A^2\beta'\tilde{V}^{2}f
T}{2\psi}+\psi\gamma=2\hat{K}+2\hat{V},
\end{eqnarray}
where
\begin{eqnarray}\label{veq}
\hat{V}=-\frac{A^2\beta'\tilde{V}^{2}f
T}{4\psi}+\frac{\psi\gamma}{2}+
\frac{\dot{\psi}\dot{T}}{2}-\frac{3M_{P}^{2}}{8}(\frac{f'}{f}^{2})\dot{T}^{2}
\end{eqnarray}
Then we can write the Friedman equations as following
\begin{eqnarray}\label{f1}
3M_{p}^{2}H^2=\rho_{m}+\rho=\rho_{m}+2\hat{K}+2\hat{V}
\end{eqnarray}
\begin{eqnarray}\label{f2}
2M_{p}^{2}\dot{H}=-\rho_{m}-\rho-P=-\rho_{m}-2\hat{K}
\end{eqnarray}
Also we obtain following relation for equation of state
\begin{eqnarray}\label{eqs}
\omega=\frac{P}{\rho}=-1+\frac{1}{1+\frac{\hat{V}}{\hat{K}}}
\end{eqnarray}
Using Eqs.(\ref{f1}), (\ref{f2}) we can write
\begin{eqnarray}\label{keq}
\hat{K}=\frac{-\rho_m}{2}-M_{p}^{2}\dot{H}
\end{eqnarray}
\begin{eqnarray}\label{veq2}
\hat{V}=\frac{3M_{p}^{2}H^2}{2}+M_{p}^{2}\dot{H}
\end{eqnarray}
As in present model, the dark energy fluid does not couple to the
background fluid, the expression of the energy density of dust
matter in respect of redshift $z$ is \cite{bin}
\begin{eqnarray}\label{zeq1}
\rho_{m}=3M_{p}^{2}H_{0}^2\Omega_{m0}(1+z)^{3}
\end{eqnarray}
where $\Omega_{m0}$ is the ratio density parameter of matter fluid
and the subscript $0$ indicates the present value of the
corresponding quantity. Using the following relation
\begin{eqnarray}\label{zeq2}
\frac{d}{dt}=-H(1+z)\frac{d}{dz},
\end{eqnarray}
one can rewrite $\hat{K}$, $\hat{V}$ as following
\begin{eqnarray}\label{keq2}
\hat{K}=\frac{-3}{2}M_{p}^{2}H_{0}^2\Omega_{m0}(1+z)^{3}+\frac{1}{2}M_{p}^{2}H_{0}^2(1+z)r'
\end{eqnarray}
\begin{eqnarray}\label{veq3}
\hat{V}=\frac{3}{2}M_{p}^{2}H_{0}^2r-\frac{1}{2}M_{p}^{2}H_{0}^2(1+z)r'
\end{eqnarray}
where
\begin{eqnarray}\label{req}
r=\frac{H^2}{H_{0}^2}
\end{eqnarray}
We obtain tachyon field in term of $z$ from Eqs. (\ref{eq4}),
(\ref{veq}), (\ref{zeq2}) and (\ref{req}) as,
\begin{eqnarray}\label{ddott1}
rH_0^2(1+z)^2T''-2rH_0^2(1+z)T'+\frac{1}{2}r'H_0^2(1+z)^2T'\nonumber\\
-\frac{2\left[(\frac{ff''+\beta'f'}{f^{2}})rH_0^2(1+z)^2TT'^{2}+2
(\alpha'-2\beta'\frac{f'}{f}T)rH_0^2(1+z)T'\right]}{1+\frac{2\alpha'}{\beta'}-3\frac{f'}{f}T-
\frac{3M_{P}^{2}}{2}(\frac{f'}{f})^{2}\frac{T}{\psi}}=0,
\end{eqnarray}
The evolution  Now using Eq.(\ref{veq})we have
\begin{eqnarray}\label{veq4}
\tilde{V}^{2}=\frac{4\psi}{A^2\beta'fT}(\frac{\psi\gamma}{2}+
\frac{1}{2}rH_0^2(1+z)^2\psi'
T'-\frac{3M_{P}^{2}}{8}\frac{f'}{f}^{2}rH_0^2(1+z)^2T'^{2}-\hat{V})
\end{eqnarray}
By using Eqs.(\ref{eqs}), (\ref{keq2}),(\ref{veq3}) we obtain
following expression for equation of state and sound speed
\begin{eqnarray}\label{eqs2}
\omega(z)=\frac{P}{\rho}=\frac{(1+z)r'-3r}{3r-3\Omega_{m0}(1+z)^3}
\end{eqnarray}
\begin{eqnarray}\label{eqs2-1}
c^{2}_s=\frac{-2r'+(1+z)r''}{-9\Omega_{m0}(1+z)^2+3r'}
\end{eqnarray}
the sound speed is discussed for investigation of stability of the
model and it necessary is to be $c^{2}_s\geq0$.\\
Then we obtain following equation for $r(z)$
\begin{eqnarray}\label{req2}
r(z)=\Omega_{m0}(1+z)^{3}+(1-\Omega_{m0})
e^{3\int_{0}^{z}\frac{1+\omega(\tilde{z})}{1+\tilde{z}}}d\tilde{z}
\end{eqnarray}
Also by using Eqs.(\ref{veq2}), (\ref{veq3}), (\ref{req}) we have
following expression for deceleration parameter $q$
\begin{eqnarray}\label{qeq}
q(z)=-1-\frac{\dot{H}}{H^2}=\frac{(1+z)r'-2r}{2r}
\end{eqnarray}
\section{Parametrization}
Now we consider the three different forms of parametrization as
following  and
compare them together.\\
\textbf{Parametrization 1:}\\
First Parametrization has proposed by Chevallier and Polarski
\cite{17}and Linder \cite{18}, where the EoS of dark energy in term
of redshift $z$ is given by,
\begin{eqnarray}\label{eqs3}
\omega(z)=\omega_0+\frac{\omega_a z}{1+z}
\end{eqnarray}

\textbf{Parametrization 2:}\\
Another the EoS in term of redshift $z$ has proposed by Jassal,
Bagla and Padmanabhan \cite{19} as,
\begin{eqnarray}\label{eqs4}
\omega(z)=\omega_0+\frac{\omega_b z}{(1+z)^2}
\end{eqnarray}
\textbf{Parametrization 3:}\\
Third parametrization has proposed by Alam, Sahni and Starobinsky
\cite{20}. They take expression of $r$ in term of $z$ as followoing,
\begin{eqnarray}\label{req3}
r(z)=\Omega_{m0} (1+z)^3+A_0 + A_1 (1+z) + A_2 (1+z)^2
\end{eqnarray}
\\
By using the results of Refs.\cite{14,15,23,24}, we get coefficients
of parametrization 1 as $\Omega_{m0}=0.29$, $\omega_0=-1.07$ and
$\omega_a=0.85$, coefficients of parametrization 2 as
$\Omega_{m0}=0.28, \omega_0=-1.37$ and $\omega_b=3.39$ and
coefficients of parametrization 3 as $\Omega_{m0}=0.30, A_0=1,
A_1=-0.48$ and $A_2=0.25$.\\
The evolution of $\omega(z)$ and $q(z)$ are plotted in Fig. 1 and
Fig. 2 respectively. Also, using Eqs.(\ref{keq2}), (\ref{veq3}) and
the three parametrizations, the evolutions of $ \hat{K}(z)$ and $
\hat{V}(z)$ are shown in
Fig. 3 and Fig. 4 respectively.\\
\begin{tabular*}{2cm}{cc}
\hspace{0.25cm}\includegraphics[scale=0.65]{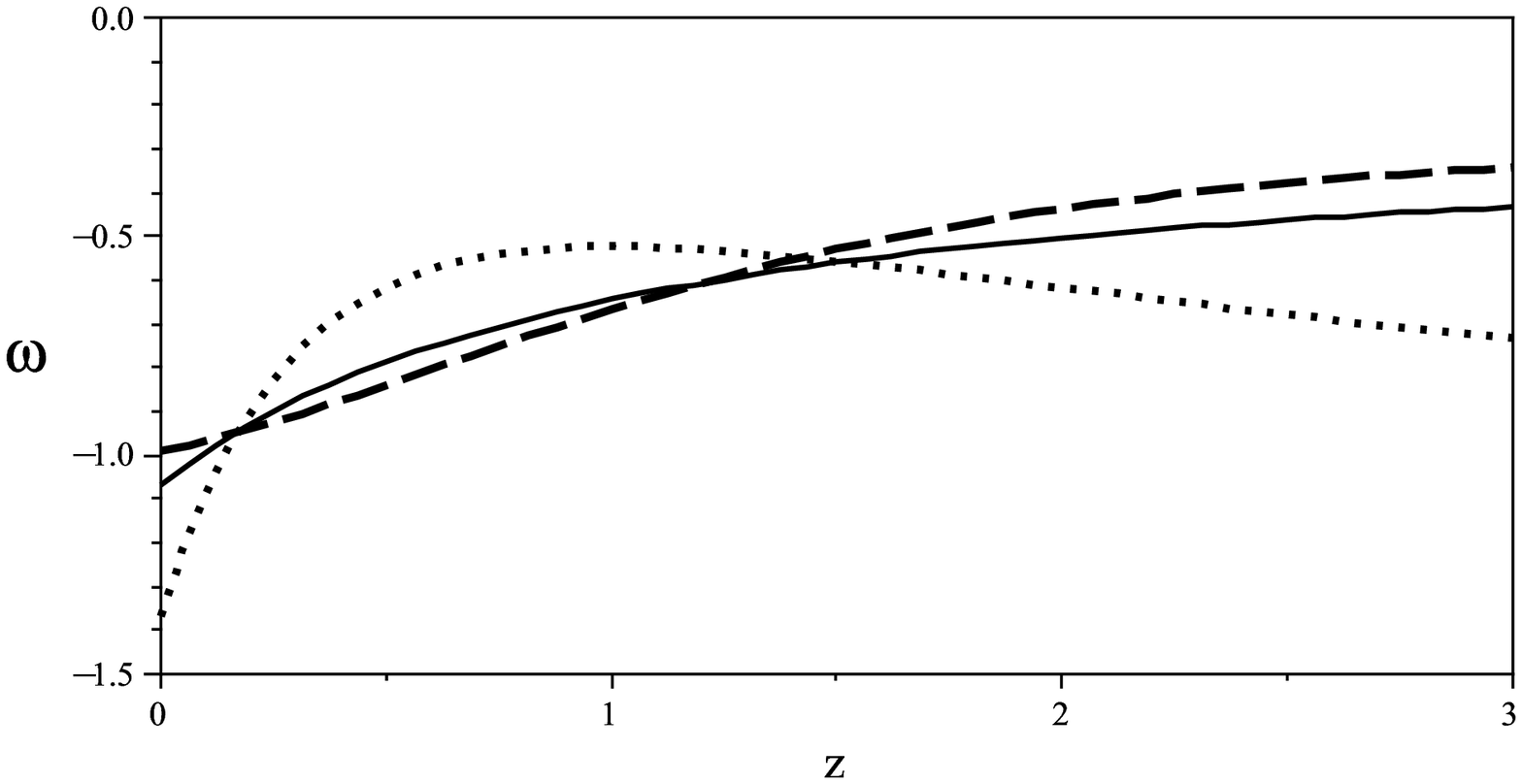}\\
\hspace{1.5cm}\textbf{Figure 1:} \,Graphs for the  EoS parameter in
respect of  redshift $z$. The solid, dot \\and dash line represent
 parametrization 1, 2 and  3
 respectively.\\
\end{tabular*}\\\\\\

\begin{tabular*}{2cm}{cc}
\hspace{0.1cm}\includegraphics[scale=0.65]{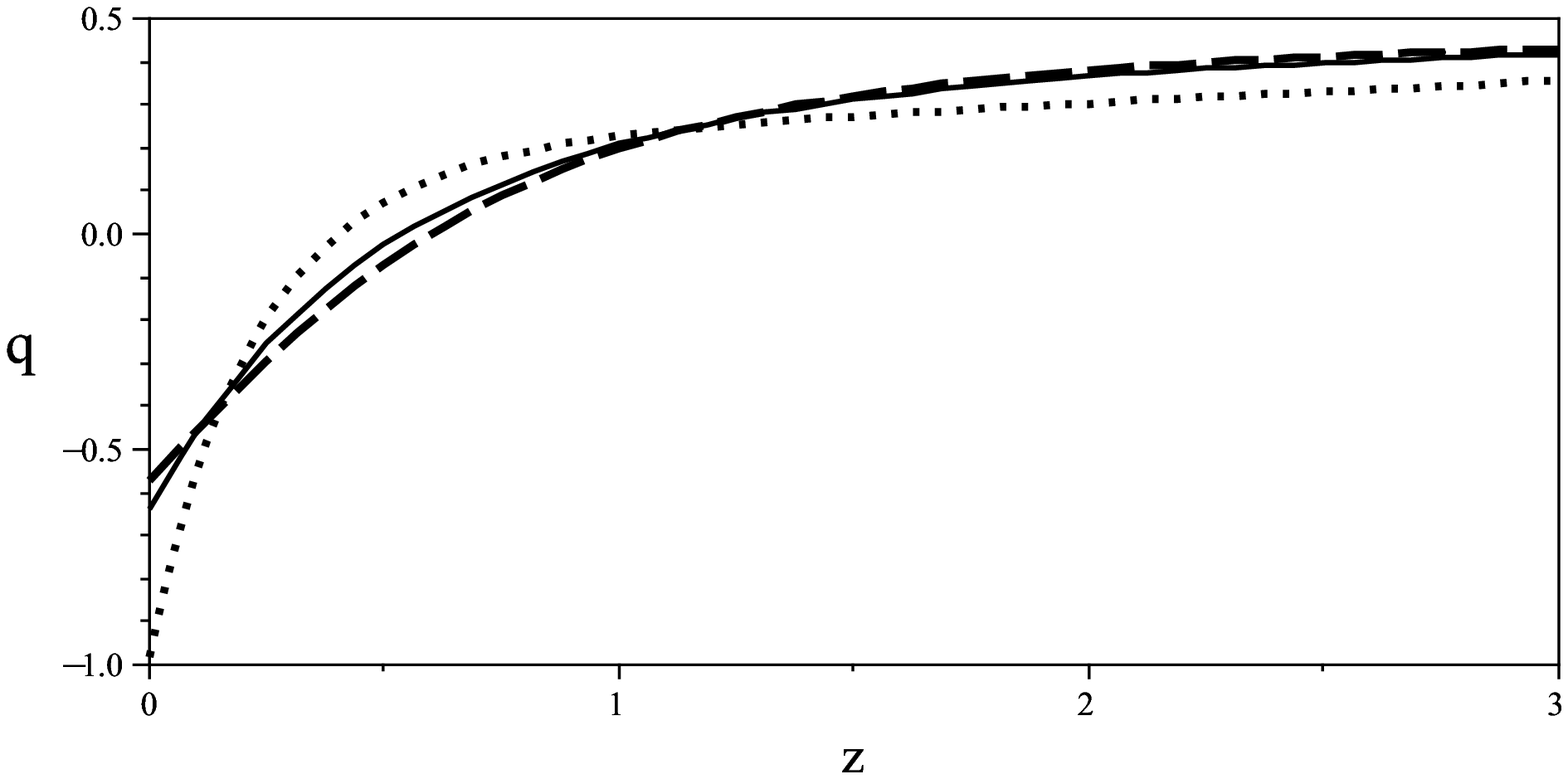}\\
\hspace{1cm}\textbf{Figure 2:} \, Graphs for the deceleration
parameter in respect of  redshift $z$. The solid,\\ dot and dash
line represent
 parametrization 1, 2 and  3
 respectively.\\
\end{tabular*}\\\\\\

\begin{tabular*}{2cm}{cc}
\hspace{0.25cm}\includegraphics[scale=0.65]{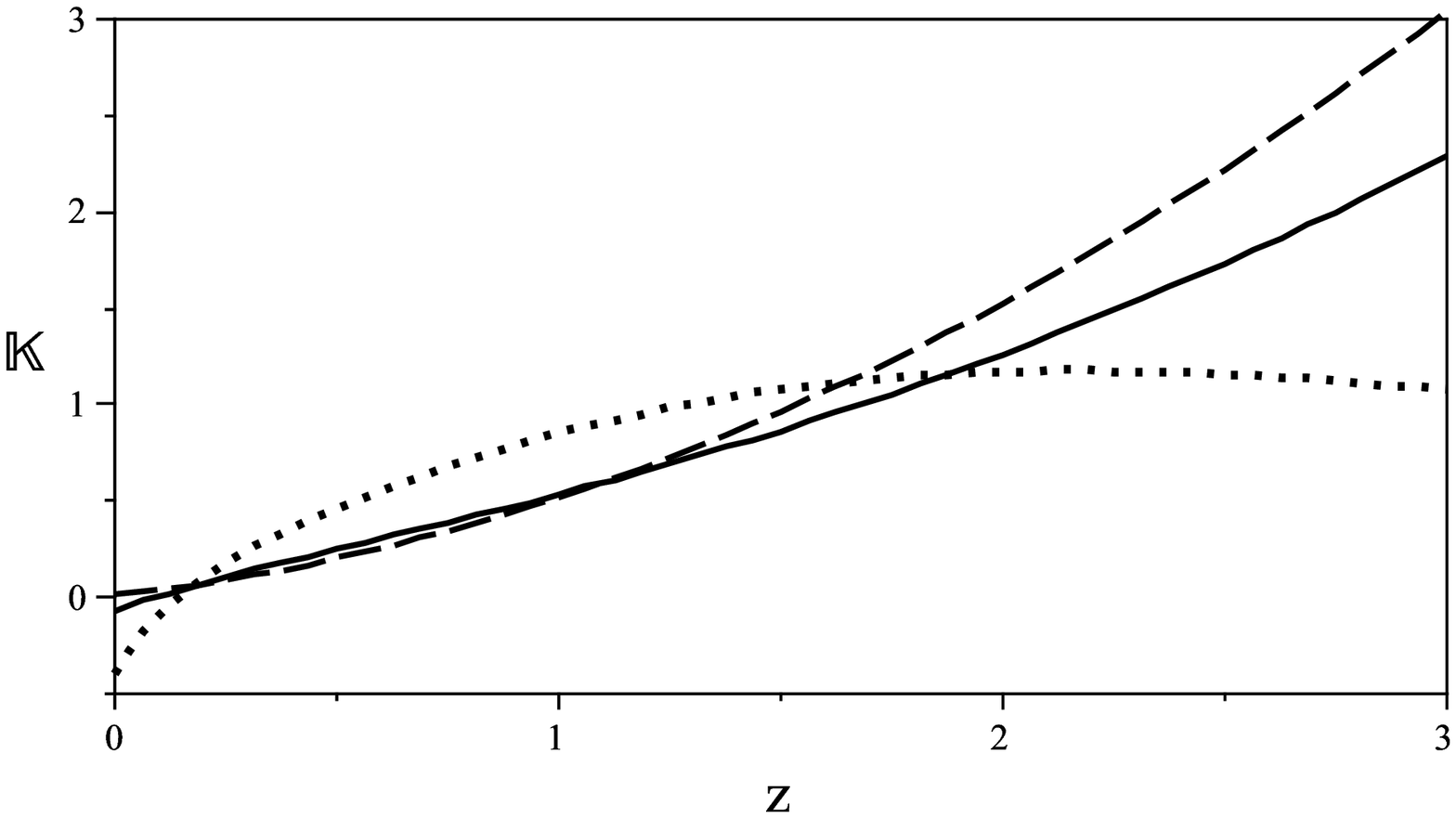}\\
\hspace{1cm}\textbf{Figure 3:} \, Graphs for the  reconstructed
$\hat{K}$ in respect of redshift $z$. The solid,\\ dot and dash line
represent
 parametrization 1, 2 and  3
 respectively.\\
\end{tabular*}\\\\\\

\begin{tabular*}{2cm}{cc}
\hspace{0.25cm}\includegraphics[scale=0.65]{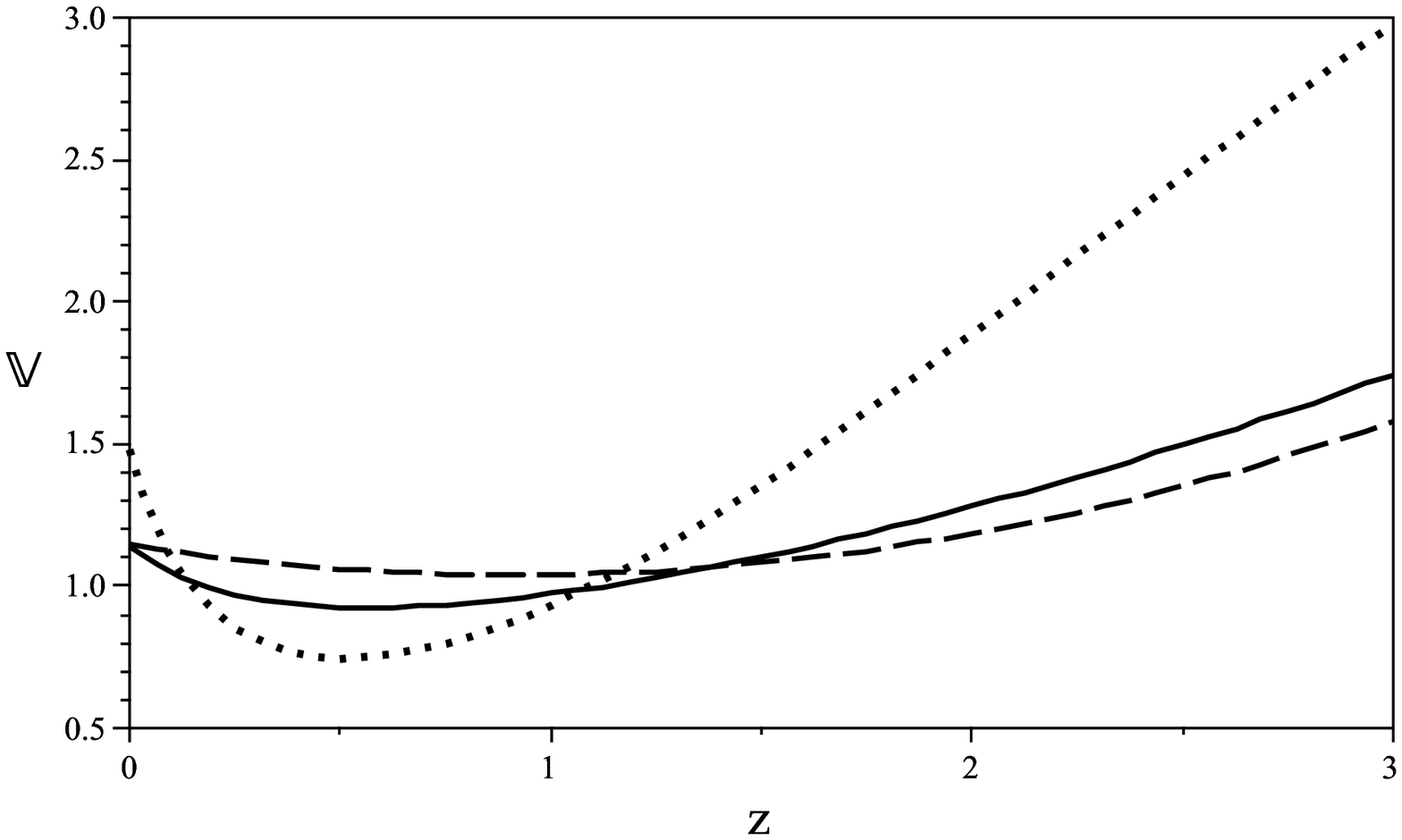}\\
\hspace{1cm}\textbf{Figure 4:} \,Graphs for the  reconstructed
$\hat{V}$ in respect of redshift $z$. The solid,\\ dot and dash line
represent
 parametrization 1, 2 and  3
 respectively.\\
\end{tabular*}\\\\\\
From Figs. (1),(2), (3) and (4), we can see parametrization 1 and 3
are same nearly and have slightly different from parametrization 2.
EoS for parametrization 1 and 3 in Fig. (1) shows to tend nearly to
$-3$, and for parametrization 2 tends nearly to $-1$. Acceleration
for all of parametrization shows to tend to the positive value. The
$\hat{K}$ and $\hat{V}$ increase for parametrization 1 and 3,
parametrization 2 increase (decrease) for the  $\hat{V}$
($\hat{K}$).\\
By using Eq. (\ref{ddott1}), we can draw $T$ with
respective to the
redshift by Runge-Kutta method in Fig. (5).\\
\begin{tabular*}{2cm}{cc}
\hspace{0.4cm}\includegraphics[scale=0.6]{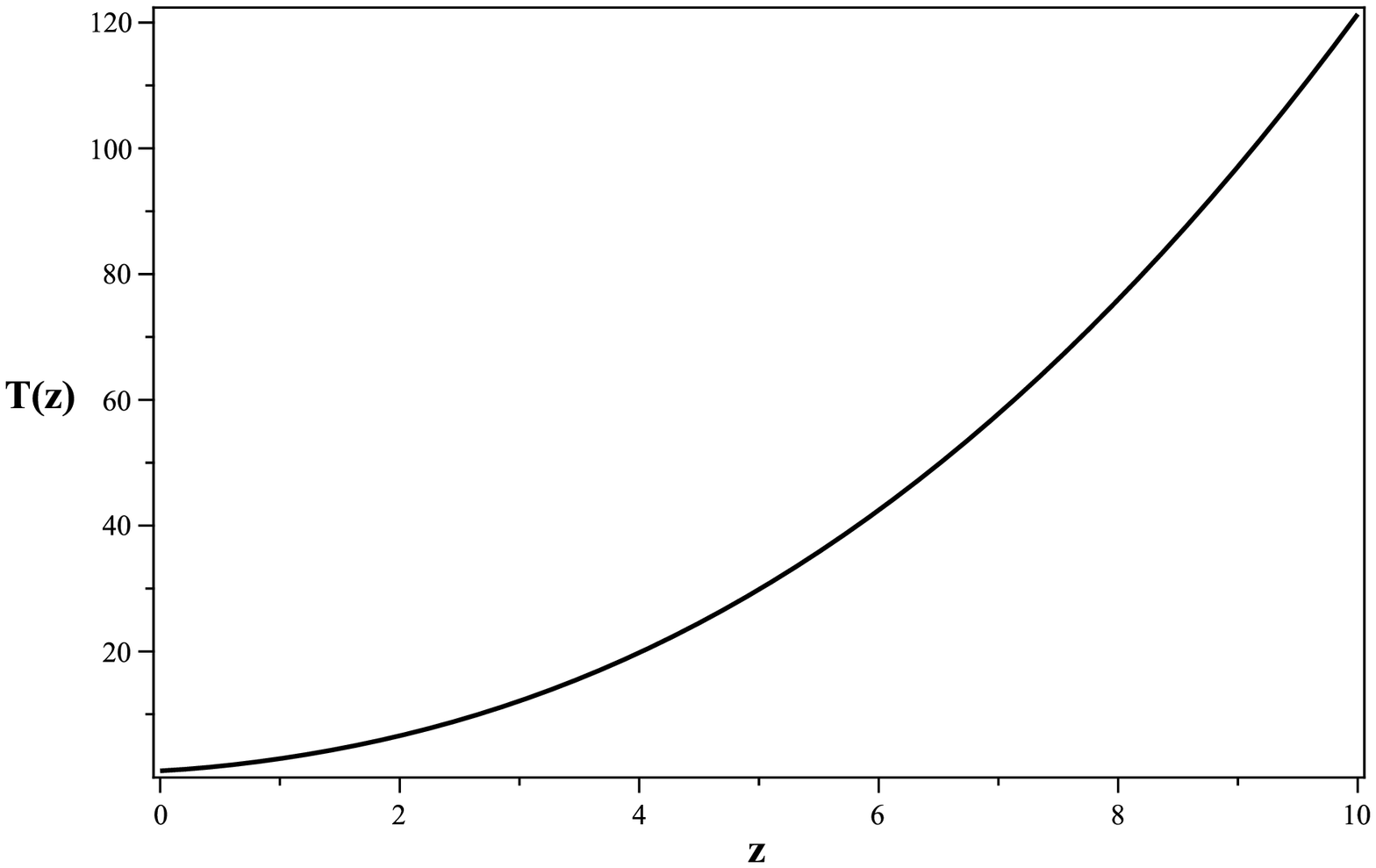}\\
\hspace{1.5cm}\textbf{Figure 5:} \,Graph for the evolution of the
tachyon scalar field in respect of\\ redshift $z$. It has calculated
by numeric procedure of
Runge-Kutta method.\\
\end{tabular*}\\\\\\
The evolution of the tachyon scalar field with respect to redshift
$z$ are same for all  parametrization 1, 2 and 3. Therefore all
parametrization give us  suitable results.  Here we note that the
second parametrization in addition to  describe the dynamics of the
tachyon scalar field is better than two others parametrization for
satisfying the EoS, one can see this point in fig Fig. 1.
Also slope of graph decrease in the early epoch.\\
Here, in order to discuss the stability of model we  use Eq.
(\ref{eqs2-1}), so we can obtain following condition,
\begin{eqnarray}\label{req5}
r(z)\geq \Omega_{m0} (1+z)^3,
\end{eqnarray}
where is accurate for three above parametrization.
\section{Conclusion}
The quintom model of dark energy \cite{{c9},{c12},{c14},{c15}} is of
new models proposed to explain the new astrophysical data, due to
transition from $\omega > -1$ to $ \omega< -1$, i.e. transition from
quintessence dominated universe to phantom dominated universe. In
this paper, we have investigated a simple method for the
reconstruction of the string-inspired quintom dark energy model with
the action (3). This  action is the same as Ref. \cite{bin} just
different to the $R f(T)$, where $f(T)$ is a function of the tachyon
$T$ and corresponds to the non-minimal coupling factor. Our aim was
to see whether the non-minimal coupling  can actually reproduce
required values of cosmological observables, such as evolution of
equation of state and the deceleration parameter in respect to the
redshift $z$. Our result for effective kinetic energy $\hat{K}$ is
exactly the same as Ref. \cite{bin}, but our result for effective
potential energy $\hat{V}$, is different with \cite{bin}. However,
our results for equation of state and the deceleration parameter in
respect to the redshift $z$, are exactly similar to what have been
obtained by the authors of \cite{bin}. Finally, we have
reconstructed our model in the light of three forms of
parametrization for dynamical dark energy. In Fig. 1 we have found
that all the three forms of parametrization require a model that
permits equation of state to cross cosmological constant boundary,
$\omega=-1$, but the second parametrization in addition to describe
the dynamics of the tachyon scalar field, it is better than two
others parametrization for satisfying the EoS.

\end{document}